# Phase- coherent comparison of two optical frequency standards over 146 km using a telecommunication fiber link

Osama Terra[1], Gesine Grosche[1], Katharina Predehl[1,2], Ronald Holzwarth[2], Thomas Legero[1], Uwe Sterr[1,3], Burghard Lipphardt[1], and Harald Schnatz[1,3]

[1]Physikalisch- Technische Bundesanstalt, Bundesallee 100, 38116 Braunschweig

[2]Max Planck Institute for Quantum Optics, Hans-Kopfermann-Strasse 1, 85748 Garching

[3]Centre for Quantum Engineering and Space-Time Research (QUEST), Germany

Email: Osama.Terra@PTB.de

Fax: ++49 531 5924305    Tel : ++49 531 5924310

**Abstract:**

We have explored the performance of two "dark fibers" of a commercial telecommunication fiber link for a remote comparison of optical clocks. The two fibers, linking the Leibniz University of Hanover (LUH) with the Physikalisch-Technische Bundesanstalt (PTB) in Braunschweig, are connected in Hanover to form a total fiber length of 146 km. At PTB the performance of an optical frequency standard operating at 456 THz was imprinted to a cw transfer laser at 194 THz, and its frequency was transmitted over the fiber. In order to detect and compensate phase noise related to the optical fiber link we have built a low-noise optical fiber interferometer and investigated noise sources that affect the overall performance of the optical link. The frequency stability at the remote end has been measured using the clock laser of PTB's $Yb^+$ frequency standard operating at 344 THz.

We show that the frequency of a frequency-stabilized fiber laser can be transmitted over a total fiber length of 146 km with a relative frequency uncertainty below $1\cdot 10^{-19}$, and short term frequency instability given by the fractional Allan deviation of $\sigma_y(\tau) = 3.3\cdot 10^{-15}/(\tau/\text{s})$.

# 1. Introduction

Recently, cold atom or single ion clocks based on optical transitions have surpassed the performance of the best microwave clocks and have the potential to reach a relative uncertainty below $10^{-17}$ together with a short–term fractional frequency instability $\sigma_y$ of a few $10^{-15} \tau^{-1/2}$ [1, 2]. Such clocks are considered for a possible redefinition of the second. However, a prerequisite for a future redefinition is the ability to compare and disseminate optical frequencies at the corresponding level of uncertainty and stability of the optical clocks.

With the advent of optical frequency comb generators [3] this is essentially solved on a local scale and different clocks in a given institute can now be compared with an uncertainty of below $10^{-16}$ [1, 2, 4, 5]. Considering that today's established techniques of frequency comparisons via satellite do not reach the required stability and uncertainty [6] and that optical clocks are not yet transportable due to their complexity, the use of optical fibers for the comparison of distant optical clocks has been discussed extensively as an innovative alternative [7, 8, 9].

Among the different methods for fiber-based frequency comparisons [see 10], the use of an AM-modulated carrier frequency [11], or the direct transfer of a highly stable optical carrier [12] are the most promising methods for long-haul frequency dissemination based on existing telecommunication fiber networks at $\lambda = 1.5$ µm.

In the first case a stable rf-frequency distribution system has been developed recently demonstrating a frequency instability for an 86 km optical link at the level of $\sigma_y = 2 \cdot 10^{-18}$ for an averaging time of 1 day [13]. However, for longer links the dispersion in the fiber, signal attenuation and the linewidth of the laser become critical issues. Additionally, the down-conversion to the microwave domain compromises the superior stability of the optical clocks.

The direct transfer of a highly stable optical carrier avoids such a down-conversion of optical frequencies. Using fiber laser based femtosecond combs [14, 15] a comparison of distant clocks can be performed in the *optical* carrier domain by measuring the *ratio* between the frequency of the laser used for transmission and the local clock frequency simultaneously at the local and remote end.

A first test of this concept was performed in 2005/06 where the remote end and the local end were located in the same laboratory and the same optical clock laser was used as a reference. A laser operating in the telecommunication window at 1.5 µm was phase-stabilized to an optical clock with a fixed and precisely known frequency ratio [16]. The transmission of the frequency of the 1.5 µm laser over an urban telecommunication fiber network in Paris showed that this method allows a fractional uncertainty below $10^{-17}$ within several hours of measurement time when fluctuations of the fiber link are eliminated. [17]. Optical carrier-phase transfer is now intensively studied by several groups to enable direct comparisons between different optical frequency standards [18, 19, 20].

In this paper, we characterize a commercially available, frequency-stabilized *dark fiber*[1] with respect to phase noise and frequency stability and demonstrate the delivery of an ultra-stable optical frequency at 194 THz over a 146 km long underground fiber to another optical frequency standard as a practical application. Section 2 summarizes the experimental setup for the realization of a stable optical frequency at 1542 nm, describes the fiber link and an optical phase stabilization system that compensates phase fluctuations induced by the optical link. In section 3 the measured instability and accuracy of the frequency provided at the remote end is presented.

This particular fiber link will eventually be used for the frequency measurement of the $^{24}$Mg frequency standard at the Leibniz University of Hanover (LUH) [21].

## 2. Experimental setup

### 2.1. Overview

The characterization of a frequency standard operated at the remote end requires a reference frequency that equals or exceeds the stability of the device under test. Furthermore, the transfer of optical stability over long-haul distances requires a highly coherent optical source, otherwise the transfer laser's frequency noise will limit the achievable stability of the fiber link. Thus, one has to reduce the power spectral density of frequency fluctuations $S_\nu$ of the transfer laser (typically $\pi S_\nu < 5 \cdot 10^3$ Hz$^2$/Hz for low-noise fiber lasers), down to the <100 Hz$^2$/Hz -level. For this, one has to develop either a cavity-stabilized and drift-controlled stand-alone system [22] or one can use a frequency comb to transfer the stability of an optical clock to the laser used for the transmission.

In this paper we use the latter approach. The set-up to derive a stable optical frequency from an optical clock laser and to measure this frequency at the remote end is shown in Fig. 1.

---

[1] "Dark fiber service" is a service provided by local exchange carriers (LEC) in which the light transmitted over the fiber is provided by the customer rather than the LEC.

In our experiment the optical frequency reference is an extended cavity diode laser (ECDL) at 657 nm ($\nu_{Ca}$ = 456 THz) stabilized to a high-finesse cavity by the Pound-Drever-Hall locking technique. This cavity-stabilized laser is used to interrogate the Ca clock transition [23] and its linewidth is below 1 Hz [24]. The small residual drift of the optical cavity has been measured with respect to a microwave frequency standard.

The near-infrared fiber laser (NIR) used as transfer laser is a commercially available single-frequency distributed feedback laser at $\lambda$ = 1542 nm ($\nu_{NIR}$ = 194 THz) (Koheras ADJUSTIK) that comprises a fiber Bragg grating written in an actively doped fiber, with a free-running linewidth below 5 kHz. Its frequency can be tuned using a piezo-electric transducer (PZT). When the NIR laser is locked to the frequency standard at 456 THz via a first optical frequency comb an ultra-stable cw-signal at 194 THz is derived (see section 2.2) [16].

This signal is then transmitted over a 146 km telecommunication fiber from the Physikalisch-Technische Bundesanstalt (PTB) in Braunschweig to the computing center of the University of Hanover and back to PTB. After a full round-trip the stability of the optical carrier transmitted through the 146 km fiber was analyzed with a second optical frequency standard in a simultaneous and independent measurement. This measurement was performed using a second frequency comb and the clock laser of the Yb$^+$ frequency standard [25] at $\lambda$ = 871 nm ($\nu_{Yb}$ = 344 THz), which has a linewidth of approx. 5 Hz [26].

The two optical frequency comb generators incorporate passively mode-locked Er-doped femtosecond fiber-lasers. Details of the setup are described in ref. [14]. The combs are pre-stabilized to radio frequencies synthesized from a H-maser (PTB-H5) to keep the beat signals within the pass band of our fixed frequency filters. These pre-filters typically have a bandwidth of 5 MHz. To allow absolute frequency measurements the maser is referenced to a primary Cs fountain clock (CSF1).

Both optical clock systems with their corresponding frequency combs are located at different buildings at PTB. They are connected with 300 m long single-mode, polarization-maintaining optical fibers at wavelengths of $\lambda$ = 871 nm and 1550 nm, respectively. For the 871 nm fiber an active stabilization system has previously been implemented to compensate frequency fluctuations due to environmental perturbations.

The various individual beat frequencies, such as the beat signals of the cw-lasers with the frequency combs and the signals of the fiber stabilization loop, that are required for the characterization of the full set-up are pre-processed by tracking oscillators (TO) fast enough to follow the frequency fluctuations of the combs, lasers and fibers. The bandwidths of the TOs are adapted individually.

## 2.2. Establishing a stable frequency reference at 1.5 µm for the frequency dissemination

Fiber-laser based femtosecond combs that bridge the frequency gap between optical clocks in the visible and the cw-fiber laser at 1.5 µm introduce phase noise into the measurement process which exceeds that of an optical frequency standard by several orders of magnitude and their control bandwidth is limited to several tens of kHz. Careful analysis of the noise sources contributing to the phase noise of fiber lasers together with sophisticated locking techniques are required to reduce this noise significantly [27].

Another possibility to overcome this problem is to circumvent the noise contribution of the comb using a special technique developed by Telle et al. [28]. Using a frequency comb as a transfer oscillator allows the comparison of lasers operating in different spectral regions without tight locking of the fs comb. Appropriate combination of optical beat signals using analogue electronics, tracking filters with a bandwidth of approx. 1 MHz and a direct digital synthesizer as a divider, results in a signal of frequency $\Delta_{trans}$ that depends on the frequency fluctuations of the cw-lasers only, and is given by

$$\Delta_{trans} = (\Delta_{Ca} + 2 f_{CEO}) \cdot m_{NIR} / m_{Ca} - (\Delta_{NIR} + f_{CEO}) = \nu_{Ca} \cdot m_{NIR} / m_{Ca} - \nu_{NIR} \quad \text{Eq. 1},$$

where $m_{NIR}$ and $m_{Ca}$ are the corresponding mode numbers of the frequency comb generating the beat signals $\Delta_{Ca}$ at 657 nm and $\Delta_{NIR}$ at 1542 nm with the cw lasers and $f_{CEO}$ being the carrier envelope offset frequency of the fs-frequency comb. $\Delta_{trans}(t)$ can be considered as a *virtual* beat frequency between a fractional multiple of the Ca clock laser frequency $\nu_{ca}$ and the frequency of the tunable laser $\nu_{NIR}$.

Locking the phase of the virtual beat signal to that of a reference oscillator using a digital phase-locked loop (dPLL) the frequency of the NIR laser can be stabilized by feedback of a control signal to a piezo-electric transducer (PZT). The control bandwidth of the loop is limited to approx. 10 kHz by a first PZT resonance at about 30 kHz. In this way we synthesize an ultra-precise and extremely stable optical reference frequency in the telecommunications window around 1.5 µm. More details can be found in ref. [16].

## 2.3. Description of the fiber link

The link to the LUH is part of a larger fiber network currently being established. This network will eventually connect optical clocks at PTB with those at the Institute of Quantum Optics (IQO) in Hanover, and the Max Planck Institutes in Erlangen (Institute of Optics, Information and Photonics, IOIP) and Garching (Max-Planck-Institute for Quantum Optics, MPQ). The fiber route was established in collaboration with the German Science Network DFN, GasLINE GmbH, and a local telecommunication provider in Braunschweig (EnBs).

A dedicated pair of dark fibers in a strand of commercially used fibers has been made available. The total fiber length from PTB to the Institute of Quantum Optics (IQO) at the university of Hanover is 73 km.

A sketch of the fiber link is depicted in Fig. 2. In a first step, our laboratory is connected to PTB's computing center that is linked to a local network provider (EnBs). The EnBs-network allows us to connect to the wide-area network of GasLINE. The networks of gas transmission and of regional distribution companies offer significant potential for developing a wide-area, fiber-optic infrastructure. The use of this infrastructure allows us to directly connect to the computing center at LUH that is located about 400 m away from the IQO. An in-house fiber link provides access to the Mg frequency standard [21] at the IQO.

For investigating the limits of the fiber link we connected the two fibers at LUH forming a loop of a total length of 146 km placing remote and local end in the same laboratory. As both fibers are located in the same strand, they are affected by the same environmental conditions and have the same characteristics. Moreover, the selection of a fiber route in the side-strip of a gas transmission line assures that the fiber is well sheltered from environmental noise leading to large frequency fluctuations.

Commercial SMF-28 fiber according to the ITU-T G.652 standard is used to establish the full link. The measured overall attenuation of the 146 km fiber link with about 16 splices and 10 connectors is approximately -43 dB. An Optical Time Domain Reflectometer (OTDR) was used to obtain detailed information of the attenuation along the optical link. This is shown in Fig. 3 for one of the fibers.

## 2.4. Description of the stabilization scheme

In order to detect and compensate phase fluctuations caused by acoustical and thermal fluctuations along the link a fiber interferometer (Fig. 4) based on commercial fiber compo-nents was constructed similar to a scheme described by L. Ma et al. [29]. At the input and the remote end of the fiber acousto-optic modulators (AOM) are installed. A Faraday mirror (FM) reflects part of the light back towards the input. The FM rotates any state of polarization (SOP) of the incoming light by 90 degrees; consequently SOP fluctuations that occur anywhere along the fiber are to a large extent compensated in the reflected light by this fully passive technique [30], and their unwanted effects on the polarisation are neutralized at the photo detector PD1. Note, this is not the case at PD2. The acousto-optic modulator (AOM-2) at the remote end is used to discriminate the desired reflection from the Faraday mirror from the backscattered light within the fiber or reflections from connectors or splices. Moreover, it allows heterodyne detection of the phase noise at a convenient rf frequency.

At the remote end the attenuation of the optical link is partially compensated using a bidirectional Er:doped amplifier (EDFA). Being part of the loop, the EDFA phase noise is suppressed within the servo bandwidth. After a full round trip of 292 km, a first interferometer measures the phase excursions accumulated along the link by comparing the light reflected from the remote end with light directly from the laser output. Phase stabilization is achieved by locking this beat signal to a local oscillator using a digital phase locked loop (dPLL) and feed-back of the control signal to AOM-1 located at the input of the fiber. Analyzing the stabilized rf signal derived from photo detector PD1 provides an in-loop measurement of the residual phase noise.

To perform an out-of-loop measurement at the remote end a second fiber interferometer measures the phase noise of the transmitted signal by means of photo detector PD2. Here polarisation changes along the fiber link can directly affect the SNR of the beat signal. Moreover, splitting off some light by implementing an additional fiber coupler at the remote end, we derive another transfer beat between the transmitted frequency at 1542 nm and the Yb$^+$ clock laser at 871 nm using a second frequency comb. We calculated this second transfer beat using the simultaneously recorded beat signals $\Delta_{Yb}$, $\Delta_{NIR}$ and $f_{CEO}$.

The reference arm of the interferometer, the acousto-optic modulators and the photo detectors PD1 and PD2 were sheltered from air fluctuations using a simple plastic crate. Without this cover we observed small changes of the signal to noise ratio (typically a few dB) of the interferometer signals which we attribute to local polarization changes. The SNR relaxed to a constant value after approximately half an hour when the crate was installed. With this setup it was not necessary to actively control the SOP during typical operational times of a few days.

## 2.5. Characterization of the fiber link

For a full characterization of the fiber link we have analyzed the phase noise of the out-of-loop and the in-loop signals of the fiber stabilization. This analysis has been accomplished in both time- and frequency-domain. In this paper we restrict the discussion to the out-of-loop signals.

We used the rf-spectrum of the beat signals to measure the *Power Spectral Density of phase fluctuations (PSD)* in the carrier domain (single sideband phase noise $L_\phi(f)$ in units of dBc/Hz). For Fourier frequencies $f < 100$ kHz we used a digital phase comparator with calibrated output and a low-frequency Fast-Fourier-Transform spectrum analyzer to measure the double sideband phase noise $S_\phi(f)$ in rad$^2$/Hz.

For signal characterization in the time domain we used a multichannel accumulating counter with synchronous readout and zero dead-time (referenced to PTB's maser H5) to record the frequencies $f_{rep}(t)$, $f_{ceo}(t)$, of each fs-comb $\Delta_{Yb}(t)$, $\Delta_{Ca}(t)$, $\Delta_{NIR}(t)$, $\Delta_{trans}(t)$, $\nu_{PD1}(t)$ and $\nu_{PD2}(t)$. This so-called π-type counter allows one to analyze the signals by means of the *Allan standard deviation (ADEV)* [31] for integration times $\tau > 0.01$ s. More information about different algorithms to calculate frequency stability with modern counters can be found in [32].

## 3. Experimental results

The basic Doppler cancellation technique described in section 2.4 relies on suppressing the optical path length fluctuations, that lead to phase noise at the remote end using a phase-locked loop at the local end. As the length of the transmission line introduces a time delay between the signal returned from the remote end and that from the local end, this delay imposes a significant constraint to the attainable stability of a fiber link, which is discussed in section 3.3.

Moreover, fluctuations of the laser frequency and the residual phase noise of the interferometer's reference arm can further limit the performance of the transfer. The latter includes contributions due to the signal to noise ratio of the detection process as well as electronic noise in the stabilization loop and unbalanced fiber. We first characterized the frequency noise of the NIR laser ($\lambda = 1542$ nm) and measured the noise floor of our interferometers [33]. Both sets of measurements have been performed in unstabilized and stabilized modes. After-wards, we measured the phase noise of the fiber link when the NIR laser was free running or locked to the optical clock laser.

### 3.1. Laser Noise

For the unstabilized NIR laser the transfer beat signal derived from a comparison with the Ca clock laser directly reflects the phase noise of the NIR laser, because the frequency standard contributes negligible phase noise. To estimate the frequency noise of the free running NIR laser we locked the laser with an attack time of $\tau_{attack} < 0.1$ s to the Ca clock laser to compensate small drifts. The spectrum of the transfer beat shown in Fig. 5 (black-dashed curve) is dominated by flicker frequency noise $S_\phi \sim (1/f^3)$ that is common for fiber lasers. The flattening of $S_\phi$ at low frequencies is due to the low gain servo loop.

After locking the NIR laser to the Ca clock laser using the transfer beat, the phase noise curve is dominated by white phase noise with an average value of $S_\phi < 1 \cdot 10^{-6}$ rad$^2$/Hz up to a locking bandwidth of 8 kHz (red curve in Fig. 5). As demonstrated in [16], the locked NIR laser then exhibits approximately the same stability as the optical clock laser. Using Eq.1, its frequency can be calculated from the known frequency of the optical reference and the measured transfer beat.

### 3.2. Interferometer Noise Floor

In our compensation scheme we were ultimately limited by the residual noise in the fiber interferometer. We have analyzed different designs of the interferometer and optimized the interferometer for compactness and minimized uncompensated fiber leads in the setup. We then measured the interferometer's noise floor when both AOMs were directly connected with a short patch cord. The frequency stability of the out-of-loop beat signal was measured for the unstabilized and stabilized interferometer. The stabilized interferometer allows us to assess the attainable noise cancellation limit without suffering from bandwidth limitation due to the fiber delay or any excess noise. More details of the interferometer design can be found in [33].

Fig. 6 and Fig. 7 show PSD and ADEV for the out-of-loop beat signal of the unstabilized and the stabilized interferometer, respectively. Calculation of the ADEV from the phase noise data [34] shows that time and frequency domain measurements agree quantitatively.

The phase noise of the unstabilized interferometer (black-dashed curve in Fig. 6) is dominated by white frequency noise of $S_\phi(f) = 6 \cdot 10^{-5}/f^2$ Hz$^2$(rad$^2$/Hz) up to a Fourier frequency $f = 1$ kHz while above 10 kHz a white phase noise level of $S_\phi(f) = 3\cdot 10^{-11}$ (rad$^2$/Hz) is reached. This is found to be in good agreement with the measured ADEV of $\sigma_y(1s) < 2\cdot 10^{-17}$ (black dots in Fig. 7). After 10 s, environmental perturbations start to dominate the stability of the unstabilized interferometer. It should be noted that even for the unstabilized interferometer the flicker noise floor is well below $1\cdot 10^{-17}$ up to an averaging time of $\tau = 10000$ s.

Once the interferometer is stabilized by means of the dPLL fluctuations of the path length and those arising from devices that are in the loop, i.e. AOMs etc. are corrected. The phase noise is then suppressed by approx. 28 dB at 1 Hz with only a slight increase of the white phase noise level between 300 Hz and 10 kHz (red curve in Fig. 6). The stabilized interferometer is dominated by flicker phase noise for $f < 1$ kHz and white phase noise for $f > 10$ kHz; for this type of noise the ADEV depends on the effective bandwidth of the frequency counting system. Assuming a high frequency cut-off of 100 kHz due to the TO tracking the signal, we calculate an instability of $\sigma_y(\tau) < 2\cdot 10^{-17}/(\tau/s)$.

At an averaging time of $\tau = 1$ s the measured instability of the stabilized interferometers (red open circles in Fig. 7) $\sigma_y(\tau) < 2\cdot 10^{-17}$ coincides with that of the unstabilized interferometer; after 1 hour averaging the instability drops below $10^{-20}$.

To determine the electronic noise limit of our measurement capabilities, the oscillator that tracks the beat signal (TO) was locked to a rf synthesizer (blue line in Fig. 7) and its frequency was counted. A fractional frequency instability of $\sigma_y(\tau) < 1.6\cdot 10^{-17}/(\tau/s)$ was achieved which includes electronic noise due to the TO, rf-synthesizer and the frequency counting system. From these measurements we conclude that the measurement noise floor of the complete systems is $\sigma_y(1\text{ s}) \leq 5\cdot 10^{-17}$ and averages approximately as $1/\tau$ with no indication of a flicker floor up to 10 000 s.

### 3.3. Transmitting an optical frequency over a 146 km fiber

A pair of dark fibers connecting PTB and LUH was connected at LUH's computing center with a patch cord to form a 146 km fiber link. Having optical reference provider and remote user both at PTB facilitated the characterization of the phase noise $S_\phi(f)$ at the remote end using PD2. We tested the fiber link both with a free-running fiber laser and after stabilizing the laser to the Ca clock laser.

### 3.3.1. Stability and phase noise of the fiber link with the free-running NIR laser

The light that generates the error signal for the fiber noise compensation (in-loop, PD1) travels about 292 km in the optical fiber. This length is much larger than the coherence length of the free running laser of about 20 km (linewidth approx. 5 kHz). As a result the laser noise in both arms of the interferometer becomes essentially uncorrelated and it will contribute to the self-heterodyne beat signals derived from PD1 or PD2.

As discussed in [35], laser noise, $S_\phi^{laser}(f)$, contributing to the interferometer signal (PD1 at the local end), $S_\phi^{self-het}$, used for the stabilization of the fiber link is given by

$$S_\phi^{self-het}(f) = 4[\sin(2\pi f nL/c)]^2 \cdot S_\phi^{laser}(f) \qquad \text{Eq. 2.}$$

For Fourier frequencies $f < (c/n\,L)$, $n = 1.468$, and $L = 146$ km this can be approximated by $S_\phi^{self-het}(f) = 8 \cdot 10^{-5} f^2 S_\phi^{laser}(f)$.

We have measured the noise of the interferometer signal at the remote end (PD2) for the unstabilized and stabilized link and the results for the PSD and ADEV are shown in Fig. 8 and Fig. 9, respectively. The contribution of the laser noise to the interferometer signal (Fig. 8, green open squares) at the remote end is calculated from the measured laser noise (blue-open squares, see also Fig. 5) using Eq.2 and replacing L by L/2. The flicker frequency noise ($1/f^3$) of the NIR laser is converted into flicker phase noise ($1/f$) by the interferometer.

The measured phase noise of the unstabilized fiber link (black dots) is fully dominated by this laser noise for f > 1 Hz; only for very low frequencies f < 1 Hz the noise of the optical link $S_\phi^{fiber}$ exceeds $S_{self-het}$(remote) of the free running laser. At about 15 Hz some technical noise shows up, which is also attributed to the link but is not yet identified. The corresponding ADEV shows a constant floor of $4 \cdot 10^{-14}$ for $\tau < 100$ s (Fig. 9 black dots) and some averaging for longer integration times.

When the fiber control loop is closed it correlates laser noise and fiber noise in order to minimize the total noise $S_\phi^{tot}(f)$, corresponding to a linear combination of $S_\phi^{fiber}(f)$ and $S_{self-het}(f)$. For the stabilized interferometer (red curve in Fig. 8) the total noise $S_\phi^{tot}(f)$ at PD2 and the measured ADEV of $\sigma_y(\tau) \approx 2 \cdot 10^{-14}/(\tau/\text{s})$ (red curve in Fig. 9) is shown.

However, the stability of the signal measured with PD2 should not be confused with the stability of the optical carrier frequency at the remote end that remains limited by the noise of the free-running laser as long as $S_\phi^{laser}(f) > S_\phi^{fiber}(f)$. Therefore, frequency stabilization of the transfer laser before compensating the phase noise resulting from the fiber link is essential.

As demonstrated earlier [16], stabilization of the transfer laser to an optical clock reduces the linewidth of the laser to a few Hz and this issue becomes negligible for distances discussed in this paper.

### 3.3.2. Stability and phase noise of the fiber link with stabilized NIR laser

By stabilizing the NIR laser to the Ca clock laser, as described in section 2.2, the noise of the fiber link $S_\phi^{fiber}$ can be measured without degradation by laser noise.

The phase noise introduced by the optical link is represented by the out-of-loop beat (PD2) of the interferometer. The phase noise of the unstabilized link is shown as black-dashed curve in Fig. 10. It can be approximated by $S_\Phi^{fiber}(f) = \left[\frac{100\,Hz}{f} \cdot \left(1 + \frac{f}{10\,Hz}\right)^{-2} + const\right] rad^2/Hz$ (blue line); above 100 Hz the noise decreases as $1/f^3$, while for $f < 10$ Hz the frequency dependence turns to flicker phase noise. The constant phase noise level of $\approx 10^{-5}$ rad$^2$/Hz at high frequencies is given by the noise floor of the phase detection system.

Beside the noise contributions associated with the interferometer and the transfer laser itself a fundamental limiting factor of any fiber link is related to the time delay $\tau_{delay} = nL/c$ introduced by the fiber [35]. First, it directly affects the control bandwidth of the compensation loop. Second and more severe, the delay through the link results in an imperfect cancellation of the noise at the remote end. This incomplete suppression of the one-way fiber noise due to the delay dominates other limitations at low Fourier frequencies.

For the round-trip signal used for the stabilization of the 146 km link, theoretically attainable noise suppression at a Fourier frequency $f = 1$ Hz is about 52 dB and the bandwidth of the loop is limited by $1/4\tau_{delay}$ to approx. 350 Hz. As a result, the stability at short averaging times (high frequencies), is limited (due to the limited bandwidth) by unsuppressed fiber noise on the link, and at longer averaging times by the residual phase noise of the interferometer. When the fiber stabilization loop is closed the phase noise of the link is suppressed to a white phase noise level of approximately $6 \cdot 10^{-3}$ rad$^2$/Hz. Due to the locking bandwidth of approximately 350 Hz, the technical noise at 15 Hz is suppressed only by approx. 25 dB. However, compared to the operation of the fiber link with the free-running laser the PSD of the link is improved by a factor of 100 at low frequencies.

The corresponding ADEV measurement (Fig. 11) shows a good agreement with the measured phase noise data. Note, that the instability of the unstabilized 146 km fiber link (black dots) is already at the low-$10^{-14}$-level. Closing the interferometer loop, an instability of the transmitted frequency of $\sigma_y(\tau) \approx 3.3 \cdot 10^{-15}/(\tau/\text{s})$ has been obtained (red open circles), reaching a level of $\sigma_y < 4 \cdot 10^{-19}$ after 10000 s. Note that the link instability follows a $1/\tau$-slope (red dashed line) over more than 6 orders of magnitude.

To compare the measured instability of the out-of-loop signal with the data reported in [20] for a 172 km link, we integrate the phase noise spectral density over the full bandwidth from 1 Hz up to 10 kHz. A root-mean square phase excursion of $\Delta\Phi_{rms} = 1.6$ rad is obtained, which results in a ADEV of $\sigma_y(\tau) \approx 2.3 \cdot 10^{-15}/(\tau/\text{s})$ in good agreement with the measured ADEV. Filtering with only 10 Hz bandwidth, as done in [20], would lead to $\Delta\Phi_{rms} = 0.33$ rad or $\sigma_y(\tau) \approx 5 \cdot 10^{-16}/(\tau/\text{s})$. This calculated value compares well with the instability $\sigma_y(1\ \text{s}) \approx 4 \cdot 10^{-16}$ reported in [20].

### 3.4. Accuracy of the transmitted frequency

Beside the attainable stability, any small residual frequency offset between the frequency at the local site and that at the remote end is of fundamental importance. The accuracy of the transmitted frequency was checked by comparing the measured mean value of the out-of-loop signal with the expected value. Passing both AOMs (55 MHz ($+1^{st}$ order) and 40 MHz ($-1^{st}$ order)), the frequency at the remote end should be shifted by exactly 15 MHz with respect to the local end.

Fig. 12 shows a time record of the frequency deviation $\Delta\nu$ of the transmitted signal from its nominal value of 15 MHz. The observed mean deviation of the transmitted optical carrier frequency given by the sample average of the full data set was $\Delta\nu = (-1.4 \pm 3.3)$ µHz. The statistical uncertainty of the mean was calculated from the standard deviation divided by the total number of data points, assuming that the $1/\tau$ dependence (white phase noise) continues over the full period of 44 hours. We have included the relative frequency uncertainty of the averaged frequency deviation of $1.7 \cdot 10^{-20}$ as red dot in Fig. 11. This value coincides with the estimated ADEV for an averaging time of 44 hours.

A more conservative estimate for the accuracy of the transmitted optical frequency is given by the last data point of the Allan standard deviation (see Fig. 11); after 9 hours a relative frequency instability of $1.3 \cdot 10^{-19}$ is reached.

## 3.5. Measuring the NIR laser frequency transmitted over the fiber link against an Yb$^+$ clock laser

To simulate a frequency measurement between a frequency standard at the remote end of the fiber link and the local Ca optical clock laser we use another frequency comb and the clock laser of PTB's Yb$^+$ frequency standard. This allows a simultaneous and **independent** measurement of the transmitted NIR frequency stability through the 146 km fiber.

The relative instability between the optical clock lasers of the Yb$^+$ and Ca frequency standards (shown as grey open dots in Fig. 13) is known from a direct comparison [36] and assures an upper limit of the relative instability of $4\cdot10^{-15}$ at $\tau = 0.1$ s approaching a flicker floor of $2\cdot10^{-15}$ for $1 < \tau < 100$ s. This flicker floor is due to the thermal noise of the reference cavity. Thus, for $\tau < 1$ s the Yb$^+$ clock laser can be used to rapidly verify the short-term stability of the transmitted laser frequency in a fully independent way.

As mentioned, both frequency standards are located in different buildings at PTB and are connected by 300 m long polarization-maintaining fibers. For $0.01$ s $< \tau < 1$ s the short-term instability of the presently unstabilized 1.5 µm-fiber used in double-pass (green squares in Fig. 13) is at the same level as the current limit for the 146 km link and only slightly above the instability of the clock laser for $\tau > 1$ s. Therefore we did not employ an additional stabilization loop for this fiber.

The independent measurement of the NIR laser frequency at the remote end using comb 2 and the Yb$^+$ clock laser is represented by the black dots, giving an upper limit of $\sigma_y(\tau) = 2.7\cdot10^{-15}/(\tau/\text{s})$ for the short-term frequency instability ($\tau < 1$ s) of the transmitted frequency. At short averaging times this measurement coincides with the measured instability of the 146 km fiber link (red open circles) discussed above. This demonstrates that we are essentially limited by the delay introduced by the fiber link. However, the stability of one of the best optical clocks [1] (blue dashed line) is reached already after 1 s of averaging. Within several minutes of averaging time the contribution of the fiber link to a clock comparison becomes negligible and optical clocks can be compared at the level of their systematic uncertainty.

## 4. Conclusion and future applications

We have presented our current capability to remotely compare optical frequencies in a fully phase coherent way by using a carrier frequency in the optical telecommunication window. The overall experiment involved transferring the stability of a local optical frequency standard operating at 456 THz to an optical carrier frequency of 194 THz using one frequency comb, transmitting this laser light through standard telecommunication fiber and telecommunication components over 146 km to a second optical clock. PTB's Yb$^+$ clock laser at 344 THz was used to simulate the real system and to measure the stability of the transmitted frequency with respect to this laser.

We showed that a standard telecommunication fiber of 146 km length can be used to transmit the frequency of an optical frequency standard with an instability of $\sigma_y(\tau) \leq 3.3 \cdot 10^{-15}/(\tau/s)$ for 0.01 s $< \tau <$ 30000 s and a relative frequency uncertainty of $2 \cdot 10^{-20}$. We have demonstrated the link stability to exceed the stability of the best available optical clocks within several minutes. This allows us to transfer the full performance of an optical clock to a general public user.

We will use the stabilized NIR laser to compare the frequency of an optical frequency standard at PTB with that of a frequency standard based on the $(3s^2)^1S_0$- $(3s3p)^3P_1$ intercombination transition of $^{24}$Mg that is currently improved at the IQO. A frequency measurement on a thermal atomic beam with respect to a portable Cs atomic clock (HP 5071A) or a GPS disciplined quartz has been reported earlier [21]. A further considerable improvement of the uncertainty of this frequency value is expected from measurements on laser cooled, free falling atoms or atoms trapped in an optical dipole trap **and** from the implementation of a better reference frequency as provided for instance by a frequency transmitted over the fiber link.

**Acknowledgements**

The authors would like to thank the staff members of PTB who have contributed to the results discussed in this report: C. Lisdat, M. Misera, and F. Vogt as well as to express their gratitude to the colleagues from the IQO Hanover J. Friebe, A. Pape, M. Riedmann, E. M. Rasel, T. Wübbena working on the optical Mg standard. The work was partly supported by DFG through SFB 407 and by the Centre for Quantum Engineering and Space-Time Research, QUEST.

**List of References:**

**Figure Captions:**

Fig. 1: Setup for comparison of two optical frequency standards, comprising an Ca clock laser $\nu_{Ca}$ at 456 THz, a fiber laser $\nu_{NIR}$ at 194 THz, Yb clock laser $\nu_{Yb}$ at 344 THz and two fs-frequency combs. At the remote end the transfer beat was calculated from the measured individual beat signals.

Fig. 2  Fiber route from PTB in Braunschweig to LUH in Hanover. Two dedicated dark fiber pairs in a strand of commercially used fibers have been made available.

Fig. 3:  The 73 km fiber-link attenuation measured from LUH, Hanover, using an OTDR. At Raffturm the local-area network of EnBs is connected to the wide-area network of GasLINE.

Fig. 4:  Setup for active fiber noise compensation. OC: optical circulator, AOM: acousto-optical modulator, VCO: voltage controlled oscillator, PD1: in-loop photo detector, PD2:out-of-loop photo detector, EDFA: Erbium-doped fiber amplifier, FM: Faraday rotator mirror.

Fig. 5:  Phase noise spectrum of the transfer beat between the fiber laser and the Ca clock laser for very low (black-dashed) and high gain (red) of the servo loop.

Fig. 6:  Residual out-of-loop phase noise floor of the stabilized interferometer (red) and unstabilized interferometers (black-dashed).

Fig. 7:  Relative frequency instability of the out-of-loop beat for the interferometer noise floor. Curve in black (dots ●) unstabilized, and in red (open circle ○) stabilized interferometer, the electronic noise is represented by the blue line.

Fig. 8:  Phase noise of the out-of-loop beat without stabilizing the NIR laser before (black dots ●) and after (red line) stabilizing the link. The phase noise of the free running laser is shown as blue-open squares □, and the calculated laser self-heterodyne signal (green open squares □).

Fig. 9:  Relative frequency instability of the out-of-loop beat with the unstabilized NIR laser before (black dots ●) and after (red open circles ○) stabilizing the link.

Fig. 10: Phase noise of the out-of-loop beat of the 146 km link before (black-dashed line), and after stabilizing the link (red line) when the NIR laser is locked to the Ca clock laser. Blue line: approximation of $S_\phi(f)$ to guide the eye, for details see text.

Fig. 11: Relative frequency instability of the out-of-loop beat-signal after 146 km with the NIR laser locked to the Ca clock laser for the unstabilized (black-dots ●) and stabilized link (red open circles ○) together with the uncertainty of the relative frequency deviation $\Delta\nu/\nu_{NIR}$ of the transmitted signal from its nominal value of 15 MHz (red dot ●); the red dotted line is equivalent to $\sigma_y(\tau) = 3 \cdot 10^{-15}/(\tau/s)$.

Fig. 12: Time record of the frequency deviation $\Delta\nu$ of the transmitted signal from its nominal value of 15 MHz over a measurement period of 44 hours (left); histogram of the frequency values (right).

Fig. 13: Relative frequency instability of the $Yb^+$ clock-laser compared to the Ca clock laser (grey open circles ○), of 300 m unstabilized fiber (green squares ■), of the NIR laser compared to the $Yb^+$ clock laser after 146 km fiber (black dots ●), and of the out-of-loop fiber stabilization signal (red open circles ○); the red dotted line is equivalent to $\sigma_y(\tau) = 3 \cdot 10^{-15}/(\tau/s)$; for comparison: the instability of the $Hg^+/Al^+$ clock comparison according to [1] is shown as blue dotted line.

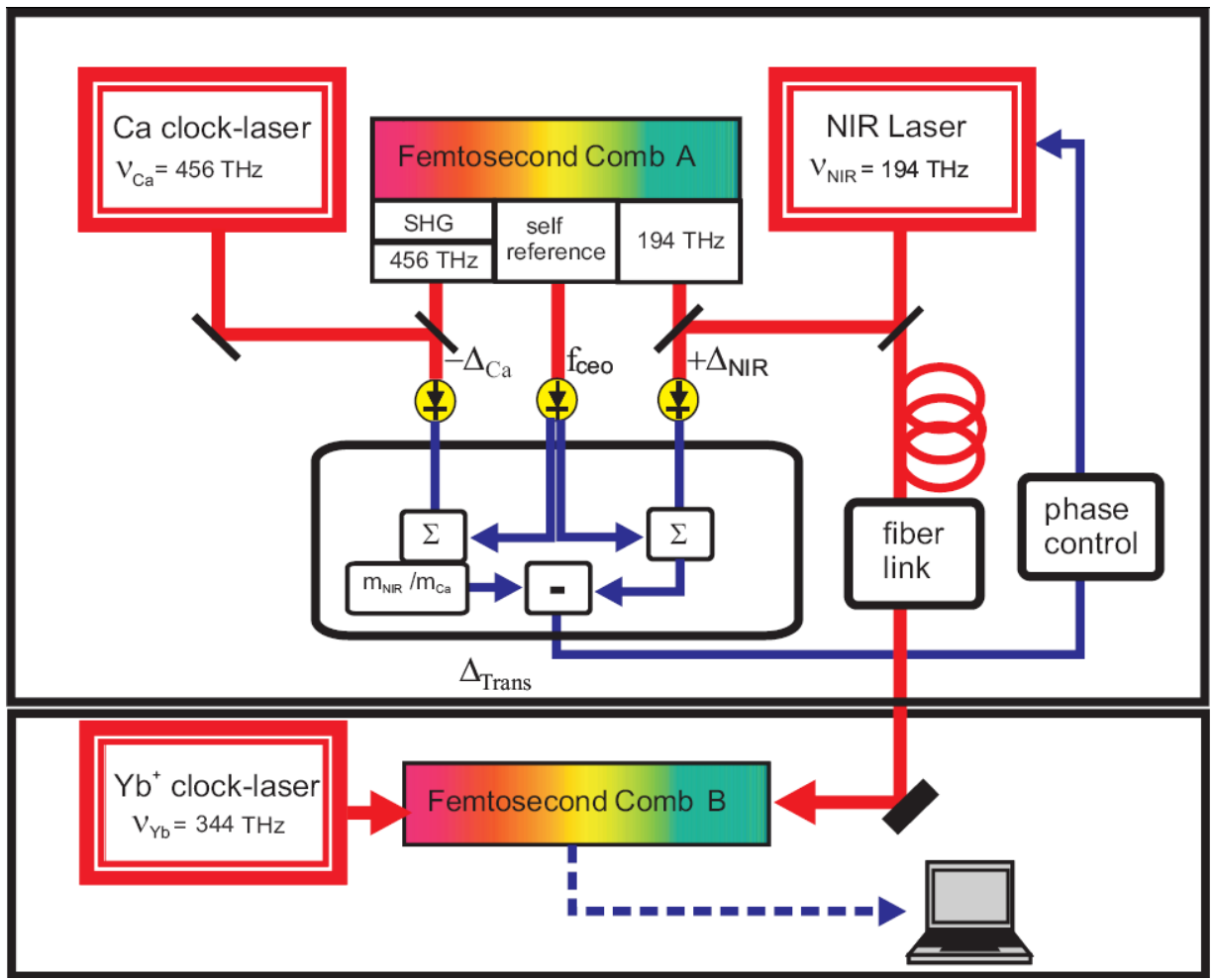

Fig. 1: Setup for comparison of two optical frequency standards, comprising an Ca clock laser $\nu_{Ca}$ at 456 THz, a fiber laser $\nu_{NIR}$ at 194 THz, a Yb$^+$ clock laser $\nu_{Yb}$ at 344 THz and two fs-frequency combs. At the remote end the transfer beat was calculated from the measured individual beat signals.

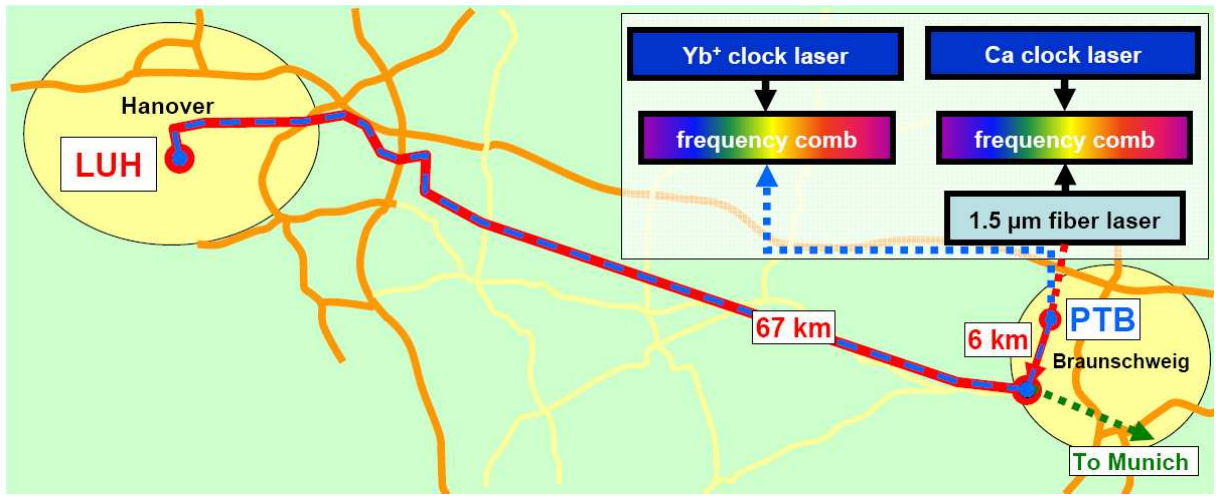

Fig. 2: Fiber route from PTB in Braunschweig to LUH in Hanover. A dedicated pair of dark fibers in a strand of commercially used fibers has been made available.

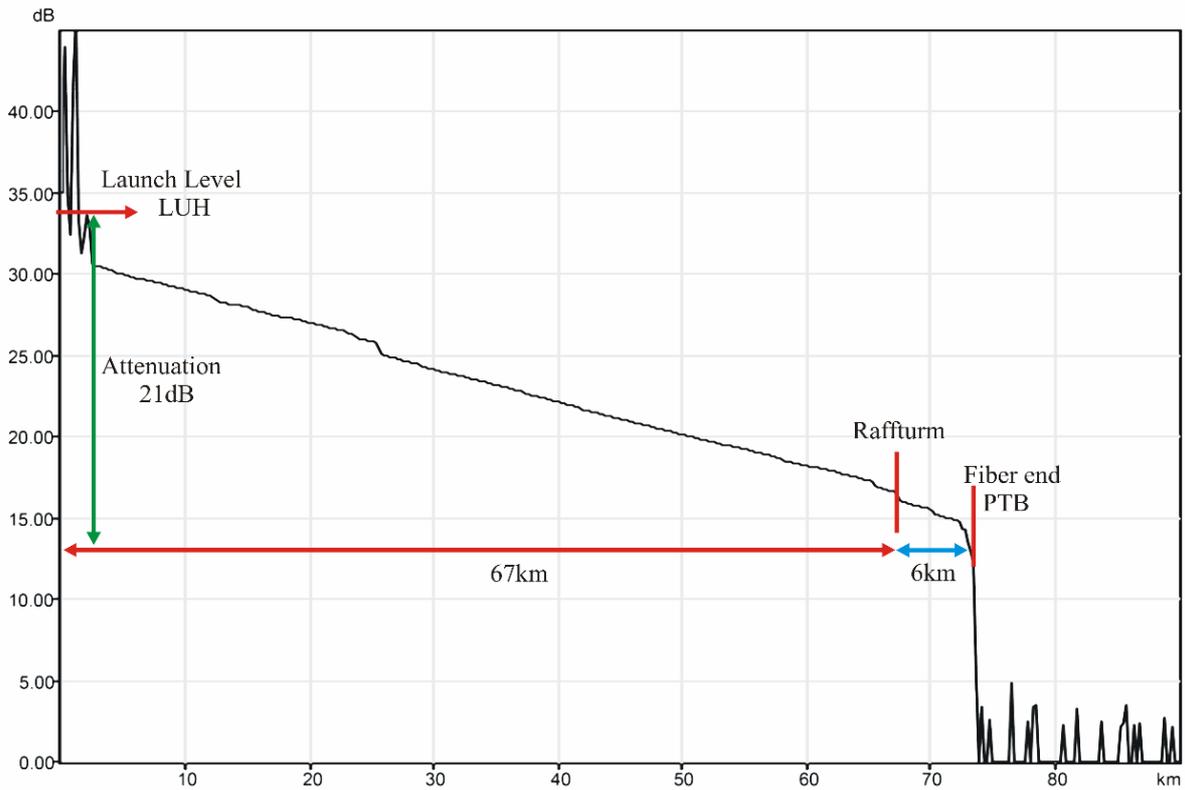

Fig. 3: The 73 km fiber-link attenuation measured from LUH, Hanover, using an OTDR. At Raffturm the local-area network of EnBs is connected to the wide-area network of GasLINE.

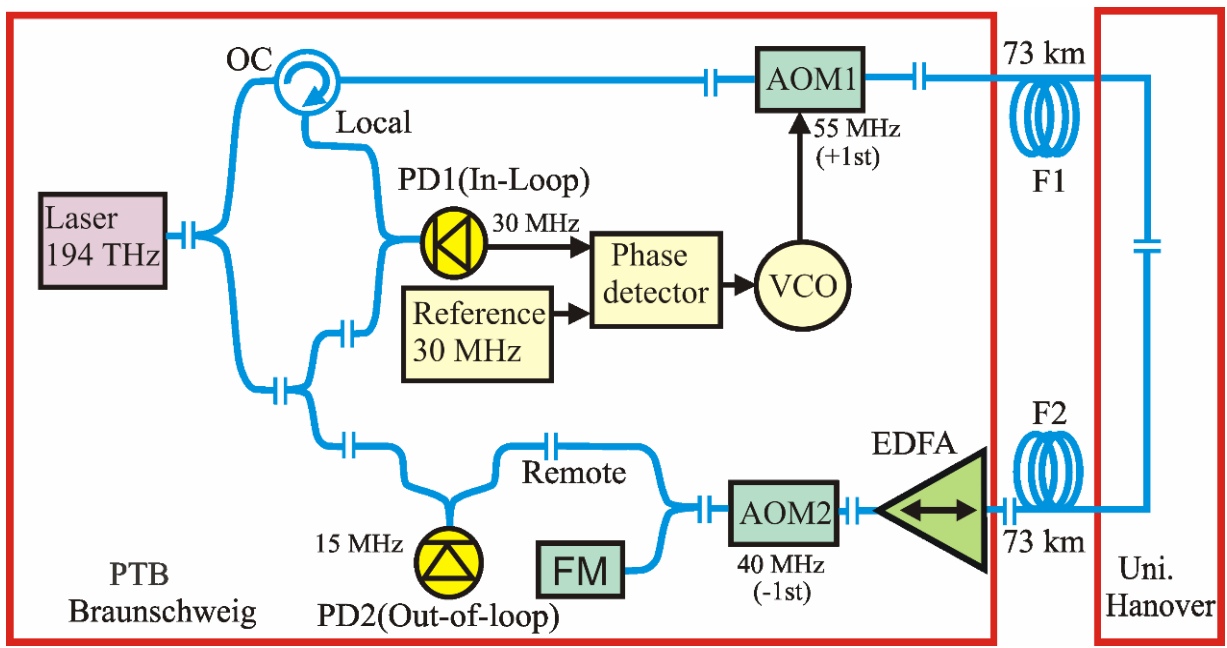

Fig. 4: Setup for active fiber noise compensation. OC: optical circulator, AOM: acousto-optical modulator, VCO: voltage controlled oscillator, PD1: in-loop photo detector, PD2: out-of-loop photo detector, EDFA: Erbium-doped fiber amplifier, FM: Faraday rotator mirror.

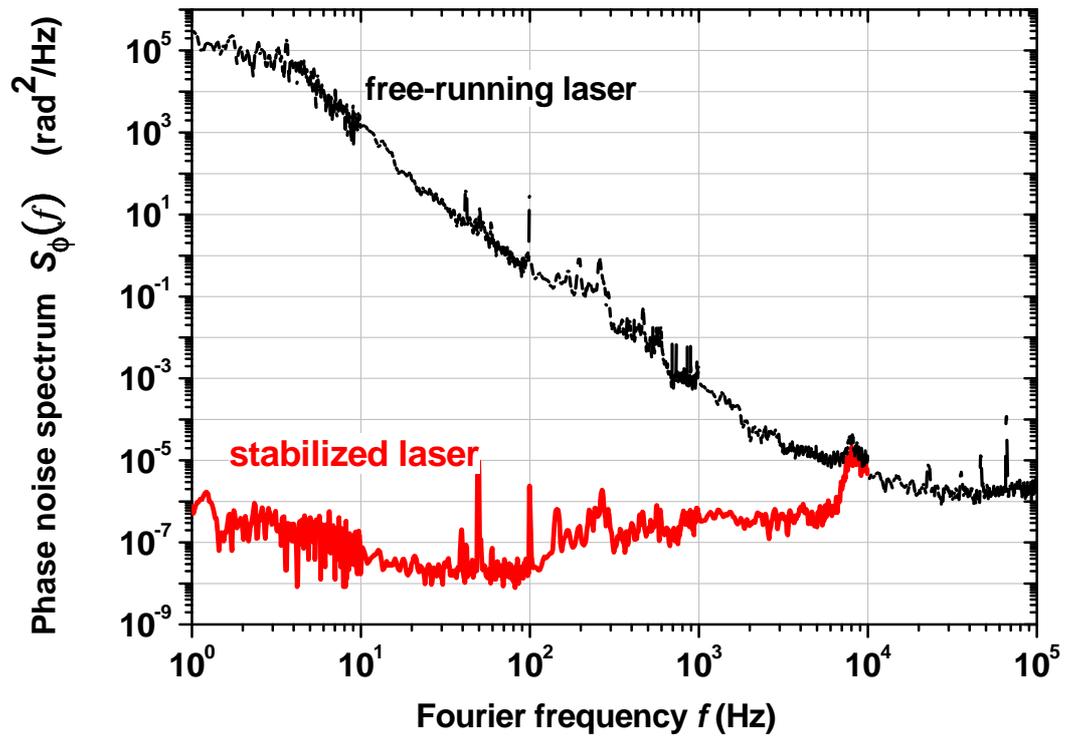

Fig. 5: Phase noise spectrum of the transfer beat between the fiber laser and the Ca clock laser for very low (black-dashed) and high gain (red) of the servo loop.

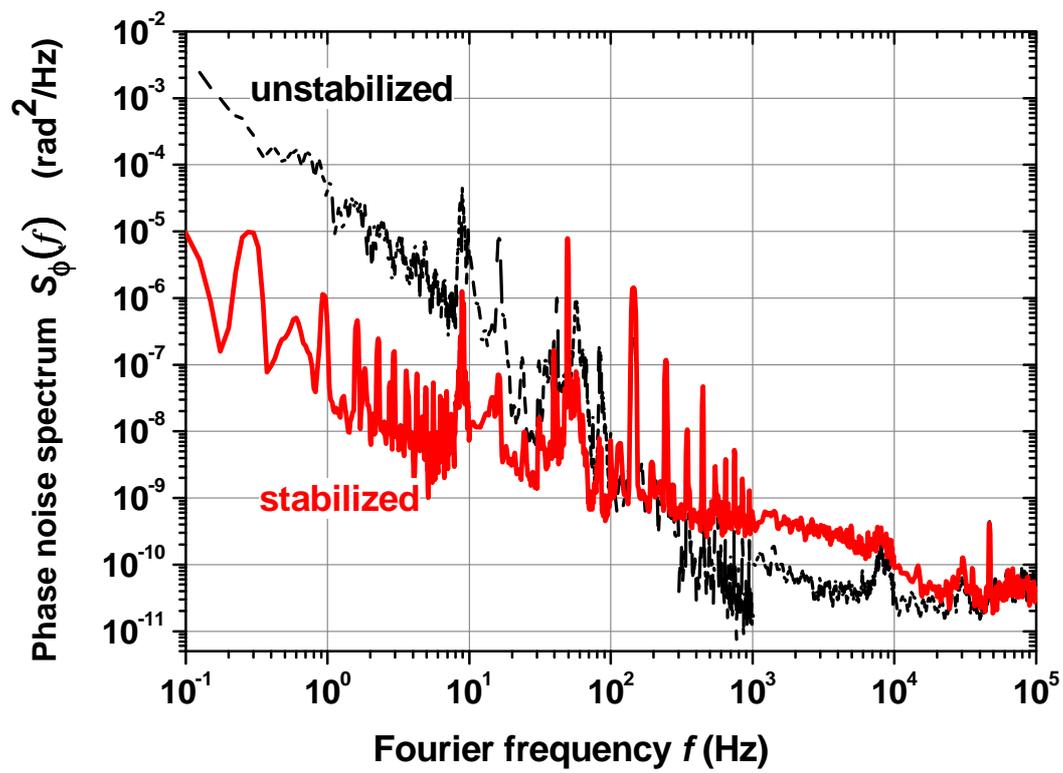

Fig. 6: Residual out-of-loop phase noise floor of the stabilized interferometer (red) and unstabilized interferometers (black-dashed).

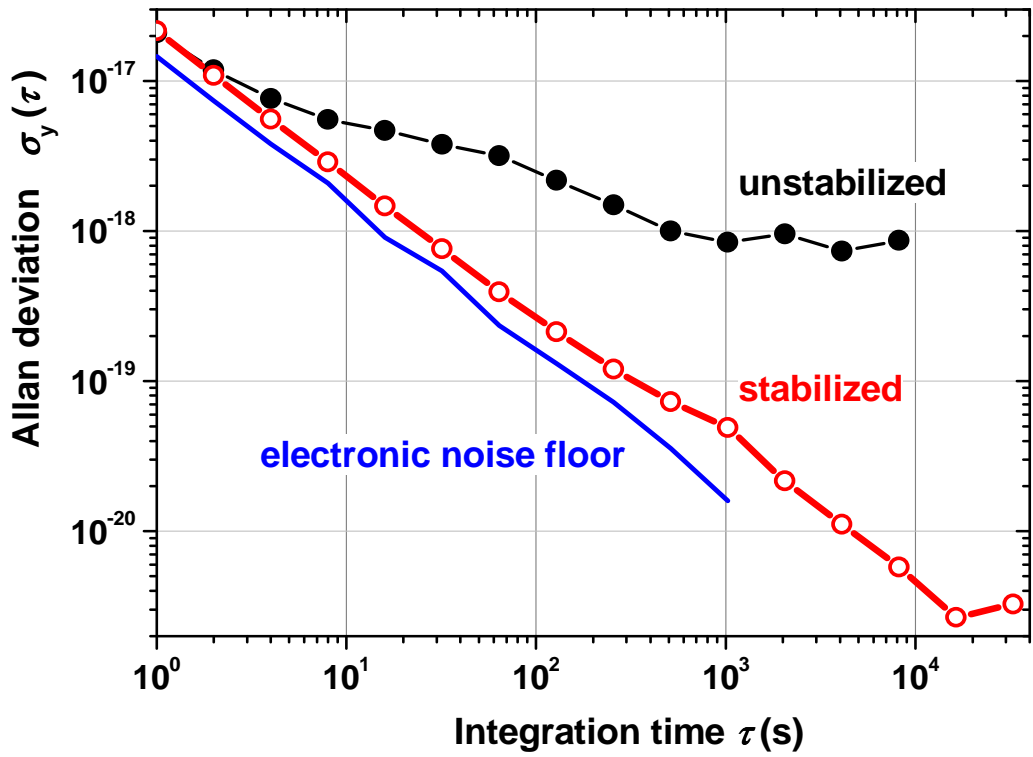

Fig. 7: Relative frequency instability of the out-of-loop beat for the interferometer noise floor. Curve in black (dots ●) unstabilized, and in red (open circle ◯) stabilized interferometer, the electronic noise is represented by the blue line.

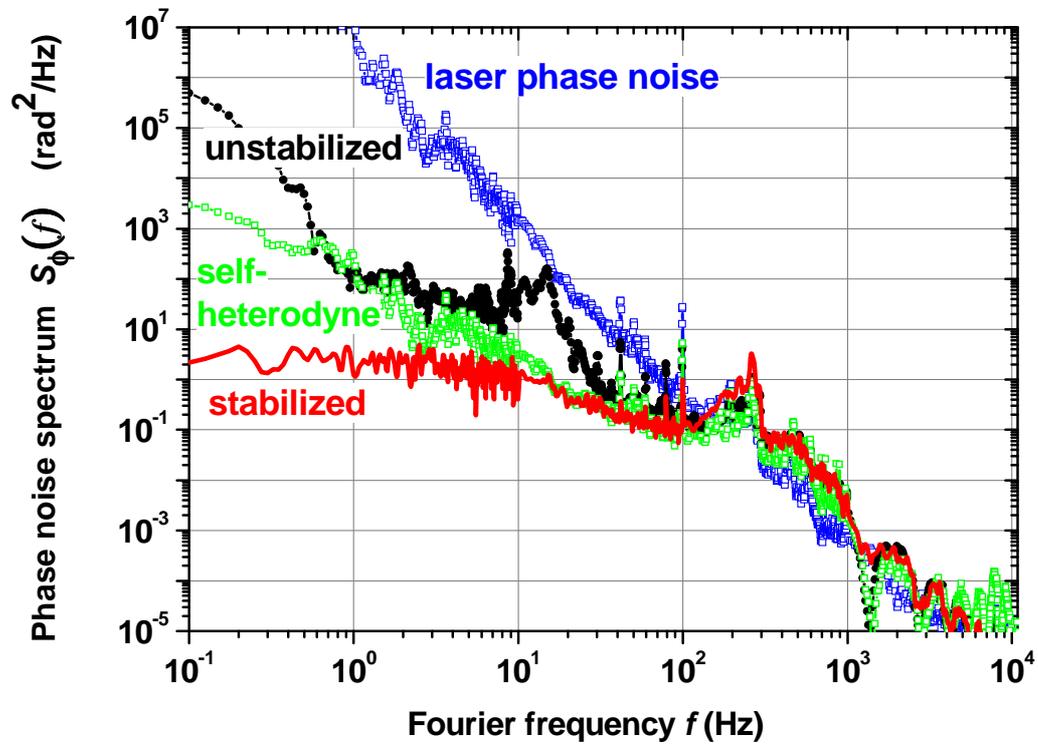

Fig. 8: Phase noise of the out-of-loop beat without stabilizing the NIR laser before (black dots ●) and after (red line) stabilizing the link. The phase noise of the free running laser is shown as blue-open squares ☐, and the calculated laser self-heterodyne signal as green open squares ☐.

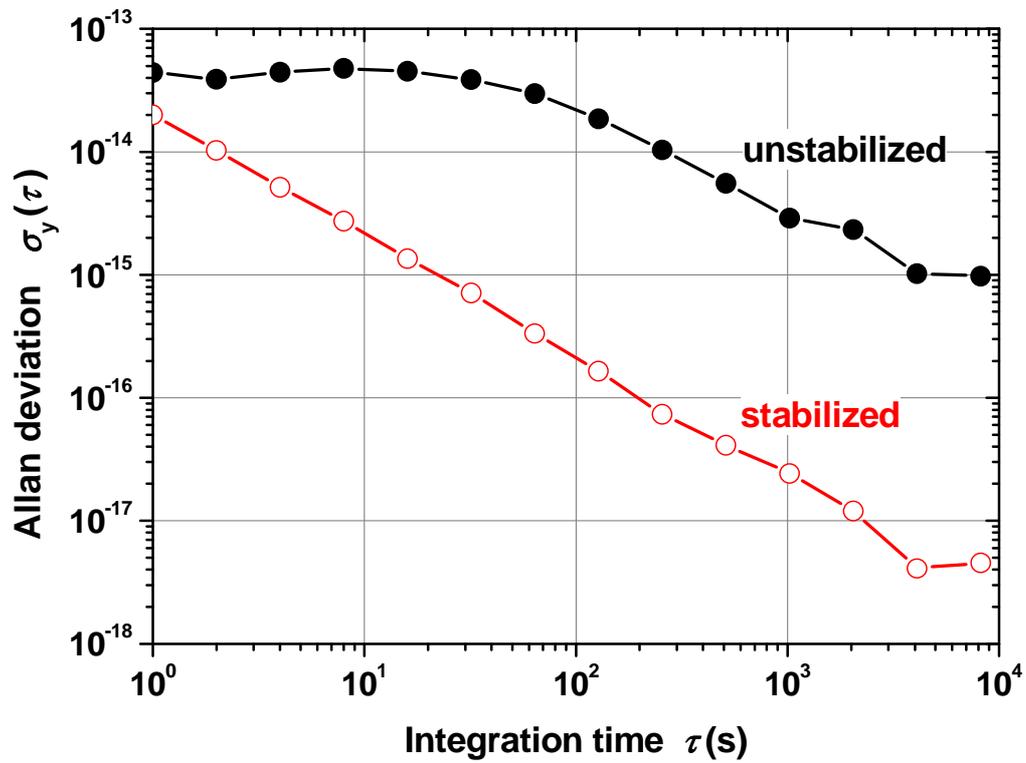

Fig. 9: Relative frequency instability of the out-of-loop beat with the unstabilized NIR laser before (black dots ●) and after (red open circles ○) stabilizing the link.

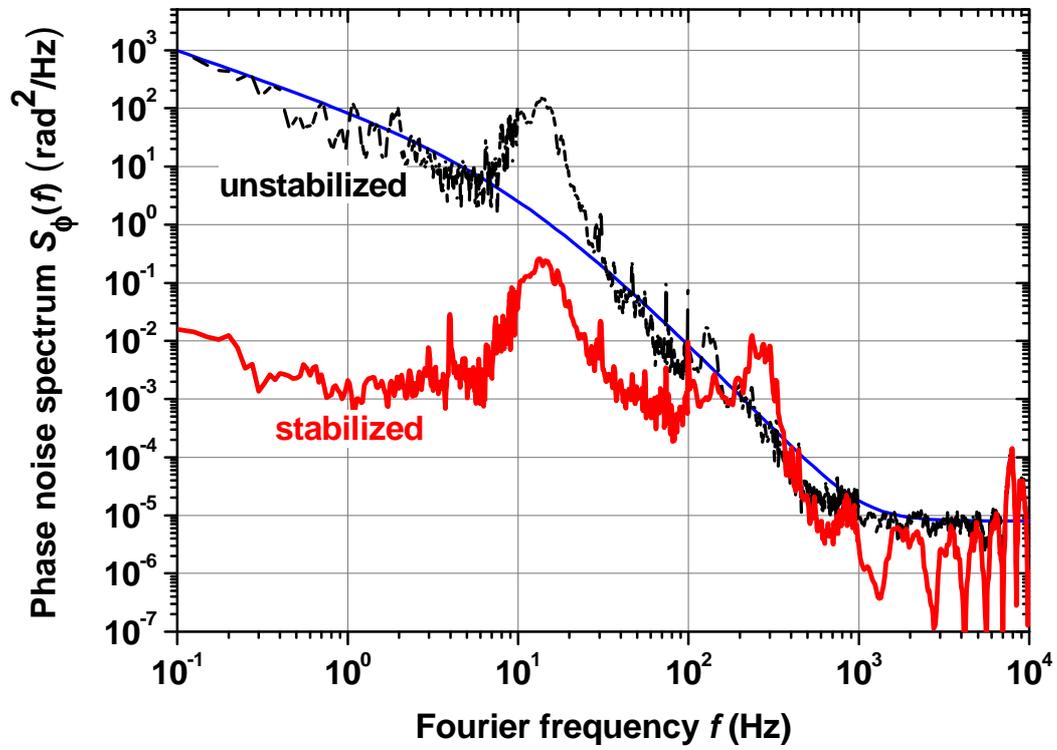

Fig. 10: Phase noise of the out-of-loop beat of the 146 km link before (black-dashed line), and after stabilizing the link (red line) when the NIR laser is locked to the Ca clock laser. Blue line: approximation of $S_\phi(f)$ to guide the eye, for details see text.

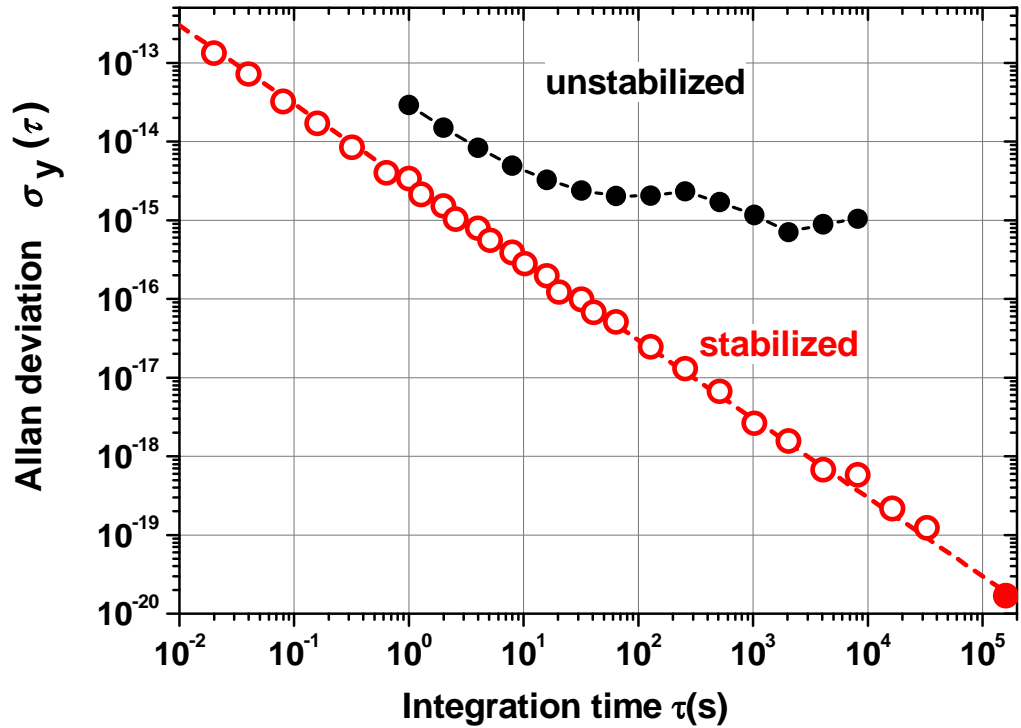

Fig. 11: Relative frequency instability of the out-of-loop beat-signal after 146 km with the NIR laser locked to the Ca clock laser for the unstabilized (black-dots ●) and stabilized link (red open circles O) together with the uncertainty of the relative frequency deviation $\Delta\nu/\nu_{NIR}$ of the transmitted signal from its nominal value of 15 MHz (red dot ●); the red dotted line is equivalent to $\sigma_y(\tau) = 3\cdot 10^{-15}/(\tau/s)$.

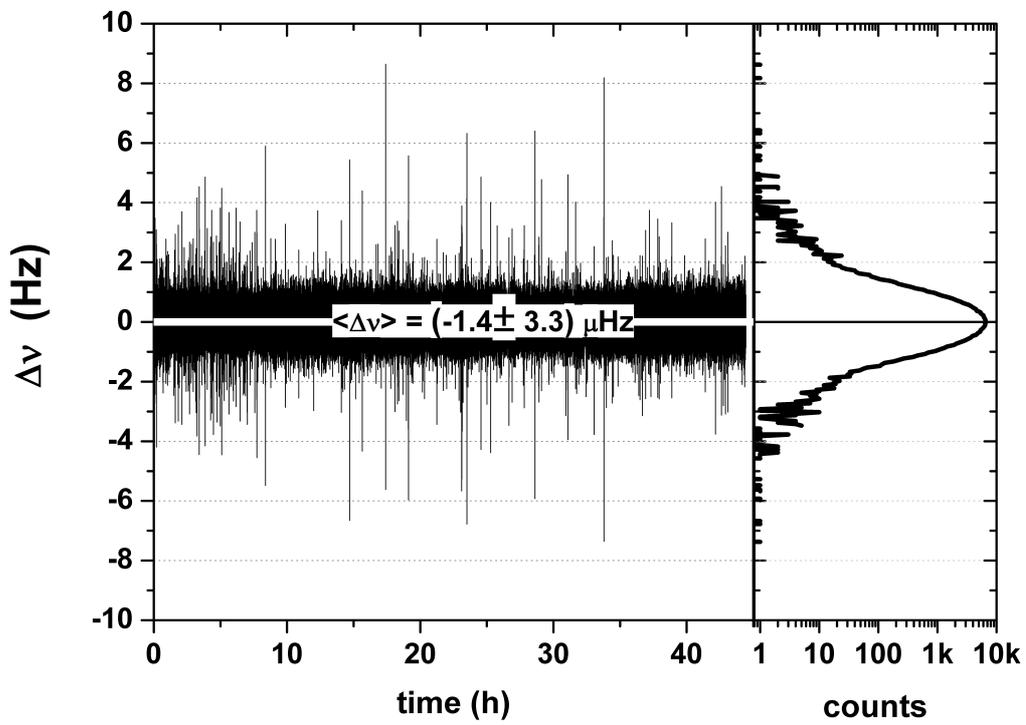

Fig. 12: Time record of the frequency deviation $\Delta \nu$ of the transmitted signal from its nominal value of 15 MHz over a measurement period of 44 hours (left); histogram of the frequency values (right).

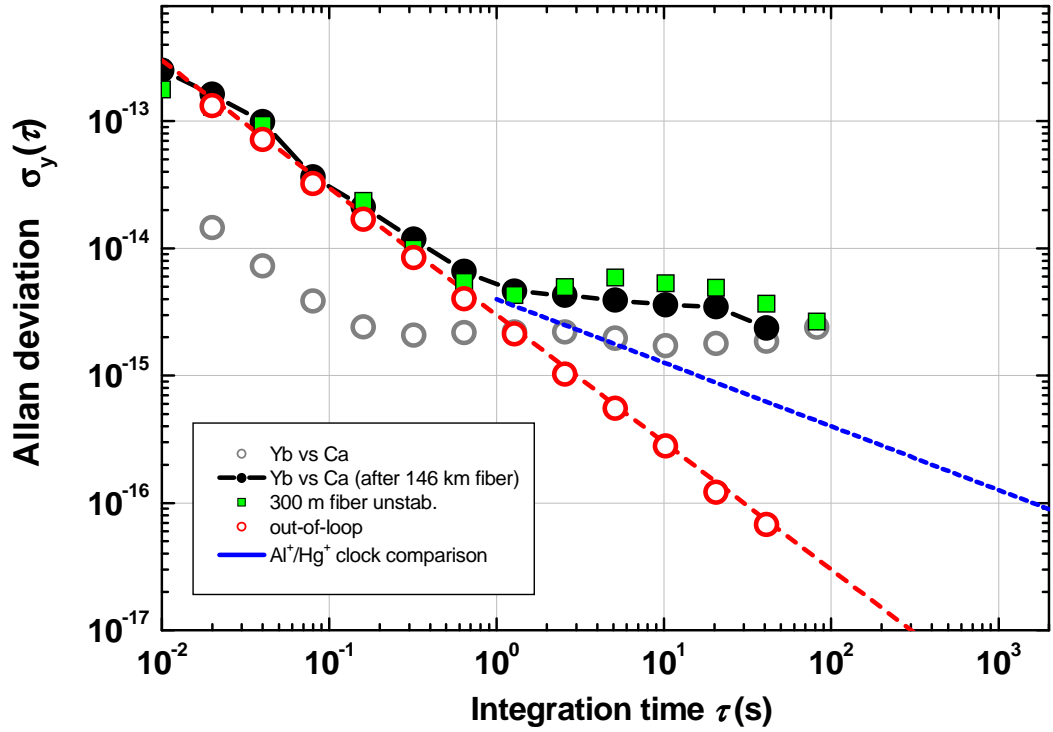

Fig. 13: Relative frequency instability of the Yb$^+$ clock-laser compared to the Ca clock laser (grey open circles ○), of 300 m unstabilized fiber (green squares ■), of the NIR laser compared to the Yb$^+$ clock laser after 146 km fiber (black dots ●), and of the out-of-loop fiber stabilization signal (red open circles O); the red dotted line is equivalent to $\sigma_y(\tau) = 3 \cdot 10^{-15}/(\tau/s)$; for comparison: the instability of the Hg$^+$/Al$^+$ clock comparison according to [1] is shown as blue dotted line.